\documentclass[useAMS,usenatbib,letterpaper]{mn2e}

\pdfoutput=1
\usepackage{amssymb} 
\usepackage[pdftex]{graphicx}

\def\btg{G\"{a}nsicke}

\def\ha{H$\alpha$}
\def\hei{He{\sc i}}
\def\heil{He{\sc i}$\;\lambda$}

\def\msun{$M_\odot$}
\def\Porb{$P_{\mbox{\tiny orb}}$}
\def\Pmin{$P_{\mbox{\tiny min}}^{\mbox{\tiny CV}}$}
\def\Psh{$P_{\mbox{\tiny sh}}$}
\def\galex{{\em GALEX}}
\def\Rsun{$R_\odot$}

\newcommand{\be}{\begin{equation}}
\newcommand{\ee}{\end{equation}}

\title[A 65-min helium-rich dwarf nova]{CSS100603:112253-111037: A helium-rich dwarf nova with a 65 minute orbital period}
\author[E. Breedt et al.]{E.~Breedt$^1$\thanks{E-mail: e.breedt@warwick.ac.uk}, B.T.~\btg$^1$, T.R.~Marsh$^1$, D.~Steeghs$^1$, A.J.~Drake$^2$, \and C.M.~Copperwheat$^1$\\
$^1$ Department of Physics, University of Warwick, Coventry, UK\\
$^2$ California Institute of Technology, 1200 E. California Blvd, CA 91225, USA
}

\begin{document}

\date{Accepted/Received}

\pagerange{\pageref{firstpage}--\pageref{lastpage}} \pubyear{2012}

\maketitle

\begin{abstract}
We present time-resolved optical spectroscopy of the dwarf nova CSS100603:112253-111037. Its optical spectrum is rich in helium, with broad, double-peaked emission lines produced in an accretion disc. 
We measure a line flux ratio \heil5876/\ha$=1.49\pm0.04$, a much higher ratio than is typically observed in dwarf novae. The orbital period, as derived from the radial velocity of the line wings, is $65.233\pm0.015$ minutes. In combination with the previously measured superhump period, this implies an extreme mass ratio of $M_2/M_1=0.017\pm0.004$.
The \ha\, and \heil\,6678 emission lines additionally have a narrow central spike, as is often seen in the spectra of AM\,CVn type stars. Comparing its properties with CVs, AM\,CVn systems and hydrogen binaries below the CV period minimum, we argue that CSS100603:112253-111037 is the first compelling example of an AM\,CVn system forming via the evolved CV channel.

With the addition of this system, evolved cataclysmic variables (CVs) now account for seven per cent of all known semi-detached white dwarf binaries with \Porb$<76$~min. Two recently discovered binaries may further increase this figure. Although the selection bias of this sample is not yet well defined, these systems support the evolved CV model as a possible formation channel for ultracompact accreting binaries.
The orbital periods of the three ultracompact hydrogen accreting binaries overlap with those of the long period AM\,CVn stars, but there are currently no known systems in the period range $67-76$ minutes. 
\end{abstract}



%

\section{Introduction}

Cataclysmic Variable stars (CVs) typically consist of a white dwarf and a near-main sequence, late-type star in a close binary orbit, with the defining characteristic that mass is transferred from the companion to the white dwarf through Roche lobe overflow. (See \citealt{warnerbook} for a detailed review of cataclysmic variables.) 

A subset of CVs show quasi-periodic brightenings of several magnitudes, known as dwarf nova outbursts.
These outbursts are thought to be the result of a thermal instability in the accretion disc which forms around the white dwarf \citep{osaki89}, and detection through outbursts remains an important method through which new CVs are discovered \citep[e.g. ][]{gaensicke05,drake09crts}. 
Some dwarf novae also show superoutbursts, which last longer than the normal outbursts and are generally brighter. During superoutbursts, the accretion disc becomes axially asymmetric due to tidal interaction with the donor star. The resulting tidal stresses and stream interaction with the outer disc are observed as a photometric modulation in the lightcurve, known as superhumps. The superhump period is closely related to the mass ratio of the system and is a good proxy for the orbital period, being typically only a few per cent longer than the orbital period \citep[e.g.][]{patterson05,kato09,gaensicke09,wood11}.

Through the process of angular momentum loss, CVs evolve from long to short orbital periods, to a minimum period \Pmin. The period minimum occurs at the point where the donor star is driven so far out of thermal equilibrium by the mass loss that it can no longer shrink rapidly enough to attain thermal equilibrium, i.e. the thermal timescale of the donor star is longer than the mass loss timescale. Further mass loss therefore requires the orbit to expand to accommodate the donor, so the system evolves back to longer periods. (See e.g. \citealt{rjw82,kolbbaraffe99}.) If the donor is partially degenerate, further mass loss will cause its radius to grow, accelerating the process. The period minimum is observed to be \Pmin$\sim80$~minutes in hydrogen-rich systems \citep{gaensicke09}. 

By definition, the donor stars in CVs fill their Roche lobes, which for a given orbital period, defines the density of the donor star.  For orbital periods much shorter than \Pmin\, the density of a Roche lobe filling star will be higher than for typical main sequence stars, which implies that the donor has to be evolved or degenerate. At present, only 42 ultracompact accreting binaries with \Porb$<76$~min are known (Section~\ref{sec:discussion}). Almost all of these belong to the small class known as AM\,CVn stars (see \citealt{solheim10} for a recent review). Their donor stars are generally deficient of hydrogen, allowing them to have very short orbital periods, observed to be in the range 5--65 minutes.
The white dwarf primary in AM\,CVn systems accrete from either another, less massive, white dwarf \citep{paczynski67,faulkner72}, from a low mass semi-degenerate helium star \citep{savonije86,ibentutukov87}, or from an evolved main-sequence star which has lost its hydrogen envelope \citep{podsiadlowski03}.  Only in HM\,Cnc, the shortest period AM\,CVn system, has hydrogen been detected in the optical spectrum \citep{roelofs10}. 
There are three different AM\,CVn formation channels, characterised by the three types of donor star seen in these systems. Their relative importance has been a long-standing question. Population synthesis models suggest that the evolved CV channel is unimportant compared to the white dwarf and helium star channels, mainly because of the long evolutionary timescales required \citep[e.g.][Yungelson et al., in prep.]{nelemans04,nelemans10}. These calculations are however sensitive to the specific form of the magnetic braking model used \citep{vandersluys05a,vandersluys05b}, and \citet{podsiadlowski03} argue that a noticable fraction of AM\,CVn stars could form from CVs with evolved donor stars.

\begin{figure*}
\centering
\rotatebox{270}{\includegraphics[width=7.5cm]{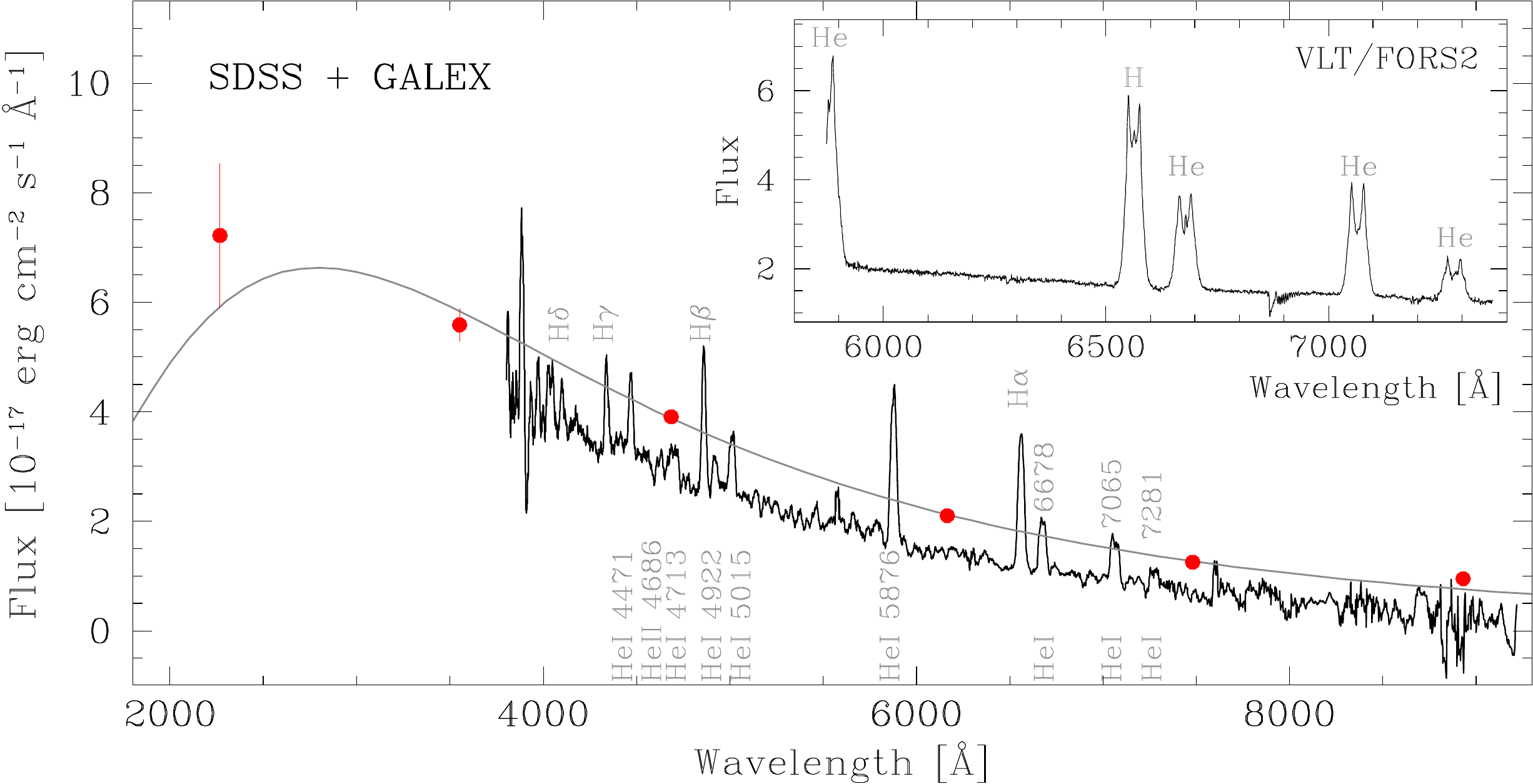}}
\caption{\label{fig:spec} {\it Main panel:}  Optical spectrum of CSS100603:112253-111037 from the SDSS, with the strongest emission lines labelled. The SDSS spectrum was smoothed by a 15 point boxcar for display purposes. The gray line overplotted on the spectrum is a blackbody of $T=10\,400$~K, fitted to the SDSS and \galex\, fluxes (red dots). The offset between the photometric and spectroscopic measurements is likely due to the variability of the source. The spectrum and the photometric measurements were corrected for Galactic extinction. {\it Inset:} Average spectrum from our VLT/FORS2 observations.}
\end{figure*}

In this paper, we present optical spectroscopy of CSS100603:112253-111037 (hereafter CSS1122-1110), a helium-dominated binary which is evolving along the CV track to become an AM\,CVn system.  It was discovered in outburst on 2010 June 3 by the Catalina Real Time Transient survey \citep[CRTS,][]{drake09crts}, at a magnitude of 14.3. It was followed up photometrically by VSNET\footnote{http://www.kusastro.kyoto-u.ac.jp/vsnet/} observers \citep{vsnet12020,vsnet12025,vsnet12026}, who reported a superhump period of \Psh=$65.434\pm0.040$~minutes \citep{kato10}.  
An optical spectrum of this system is available from the Sloan Digital Sky Survey \citep[SDSS,][]{abazajian09_sdssdr7}, shown in Figure~\ref{fig:spec}. It reveals a blue continuum dominated by \hei\, emission, along with obvious hydrogen Balmer emission lines.
Subsequently CSS1122-1110  was also included in the SDSS CV catalogue of \citet{szkody11}\footnote{Referred to as SDSS\,J112253.3-111037.6}, who also noted the unusual helium-to-hydrogen line ratios in this binary.

%
\section{Observations} \label{sec:data}

\subsection{Optical spectroscopy}

We carried out time-resolved spectroscopy of CSS1122-1110 over three nights 
in 2011 March, using the FORS2 spectrograph at the European Southern Observatory (ESO)
Very Large Telescope (VLT) at Paranal, Chile. Observing conditions 
were very good, with sub-arcsecond seeing throughout.
We used the GRIS1200R+93 grism with a 1\arcsec\, slit and 2$\times$2 binning of the 
2k$\times$4k MIT CCD. The resulting spectra cover the spectral range 5872 -- 7370\AA\,
with an average dispersion of 0.73\AA\, per binned pixel and a full width at half maximum (FWHM) resolution of 1.7\AA. 

We obtained 52 spectra in total over the three nights, covering approximately two binary orbits on the first night, two on the second night and one orbit on the last night.
All spectra were taken using an exposure time of 420 seconds, except for one exposure on the second 
night which was aborted after 177 seconds due to a temporary guiding error of the telescope. 
The spectra were reduced using the {\sc starlink}\footnote{http://starlink.jach.hawaii.edu/starlink}
packages {\sc kappa} and {\sc figaro}, and optimally extracted using {\sc pamela}\footnote{{\sc pamela} is included in the {\sc starlink} distribution `Hawaiki' and later releases.} \citep{marsh89pamela}.

The wavelength calibration is based on helium-argon-neon arc lamp exposures which were
taken during the day. We identified 25 reliable arclines over the spectral range and fitted them with a a fifth order polynomial. The root mean square (rms) of the fit residuals was 0.012\AA. 
In order to account for a small amount of flexure of the instrument, we shifted the spectra so that the strong night-sky emission line at 6300.304\AA\, appears at its known wavelength. The typical shifts required were $0.1-0.2$ pixels.

The instrumental response and flux calibration were calculated from a spectroscopic standard star observed at the start of each night as part of the ESO calibration programme. The spectra from the first and third nights were calibrated using HD49798 and the second night's spectra were calibrated against Hiltner\,600.  The spectra are not corrected for telluric absorption or slit losses, so the flux calibration is not absolute. However, the continuum slope and the relative calibration over the three nights are reliable.

\subsection{Photometry} \label{sec:phot}

\begin{figure}
\centering
\rotatebox{270}{\includegraphics[width=5.5cm]{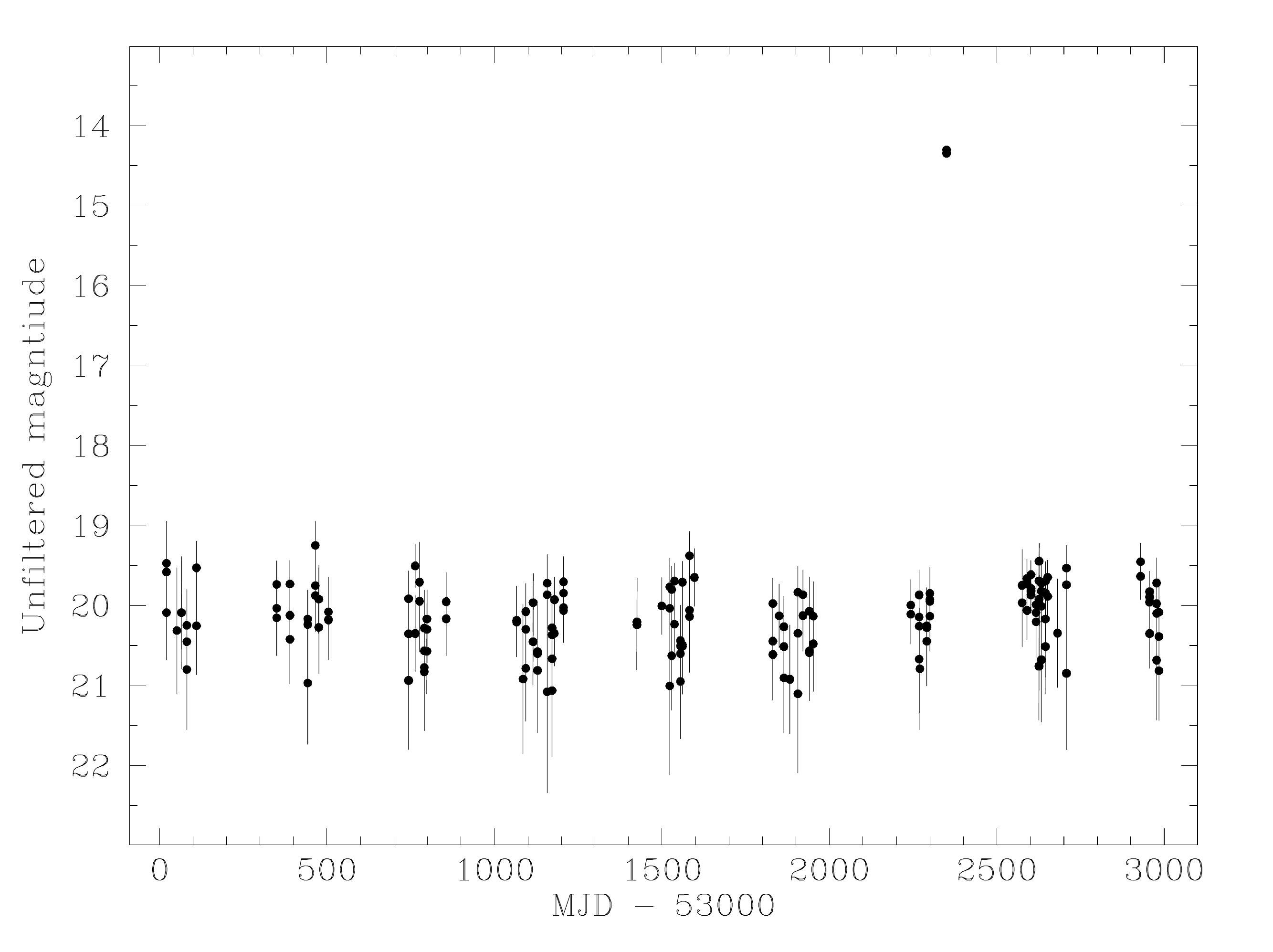}}
\caption{\label{fig:crtslc} CSS lightcurve spanning eight years of observations. The 
average quiescent magnitude is $20.2\pm0.5$ and the outburst reached a magnitude of 14.3.}
\end{figure}

The Catalina Sky Survey (CSS) has visited the field containing CSS1122-1110 256 times since the start of the survey. Our target was undetected on 82 of these images, and we rejected a further six for which the measured photometric error exceeded eight per cent (compared to the mean error of 2.3 per cent). This left 168 reliable photometric measurements. The median time between CRTS visits is 19~days, with typically 2--4 observations per visit, taken over 30--40~minutes. The lightcurve, spanning 8.1 years, is shown in Figure~\ref{fig:crtslc}. Only one outburst was seen during this time.  We measure an average unfiltered quiescent magnitude of $20.2\pm0.5$, in good agreement with the SDSS magnitudes.

We also carried out differential photometry on our VLT/FORS2 acquisition images, using {\sc iraf}\footnote{{\sc iraf} is distributed by the National Optical Astronomy Observatory (NOAO), which is operated by the Association of Universities for Research in Astronomy (AURA) under cooperative agreement with the National Science Foundation.}. The acquisition was done using the R\_SPECIAL+76 filter, a broadband filter centred on 6550\AA\, with a FWHM of 1650\AA. It therefore includes all four lines in our spectrum. The photometry showed that the object brightened by $0.15\pm0.02$ magnitudes over the three nights of observations.

\section{Analysis and results}  \label{sec:results}

\subsection{Spectral energy distribution}

The ultraviolet (UV) and optical magnitudes of CSS1122-1110, as given by the \galex\, \citep[Galaxy Evolution Explorer,][]{martin05galex} Data Release 6 and SDSS Data Release 8 \citep{sdssdr8}, are shown in Table~\ref{tab:mag}. The system is undetected in the \galex\, far-UV band.  Correcting for Galactic extinction towards the source, $E(B-V)=0.048$ \citep{schlegel98}, the photometric measurements are well described by a $T=10\,400\pm300$~K blackbody spectrum, shown as a gray line in Figure~\ref{fig:spec}. The flux difference between the photometric and spectroscopic SDSS measurements is due to the variability of the source.
A lower limit on the distance to the system can be set by assuming that the continuum emission originates from the white dwarf only, with no disc contribution. Assuming a typical white dwarf radius of 0.01\Rsun, this gives $d\geq345$~pc.

\begin{table} 
  \centering
  \caption{\label{tab:mag} Spectral energy distribution.}
  \begin{tabular}{lccc} \hline
     Band     & $\lambda_{\mbox{eff}}$\,(\AA) & Magnitude & Error \\
    \hline\hline
     \galex\, NUV & 2267 & 21.65 & 0.22 \\
      SDSS-$u$    & 3551 & 20.71 & 0.06 \\
      SDSS-$g$    & 4686 & 20.44 & 0.02 \\
      SDSS-$r$    & 6165 & 20.46 & 0.03 \\
      SDSS-$i$    & 7481 & 20.57 & 0.04 \\
      SDSS-$z$    & 8931 & 20.46 & 0.12 \\
     \hline 
  \end{tabular}
\end{table}

\subsection{Emission line profiles}

\begin{figure*}
\centering
\rotatebox{0}{\includegraphics[width=12.5cm]{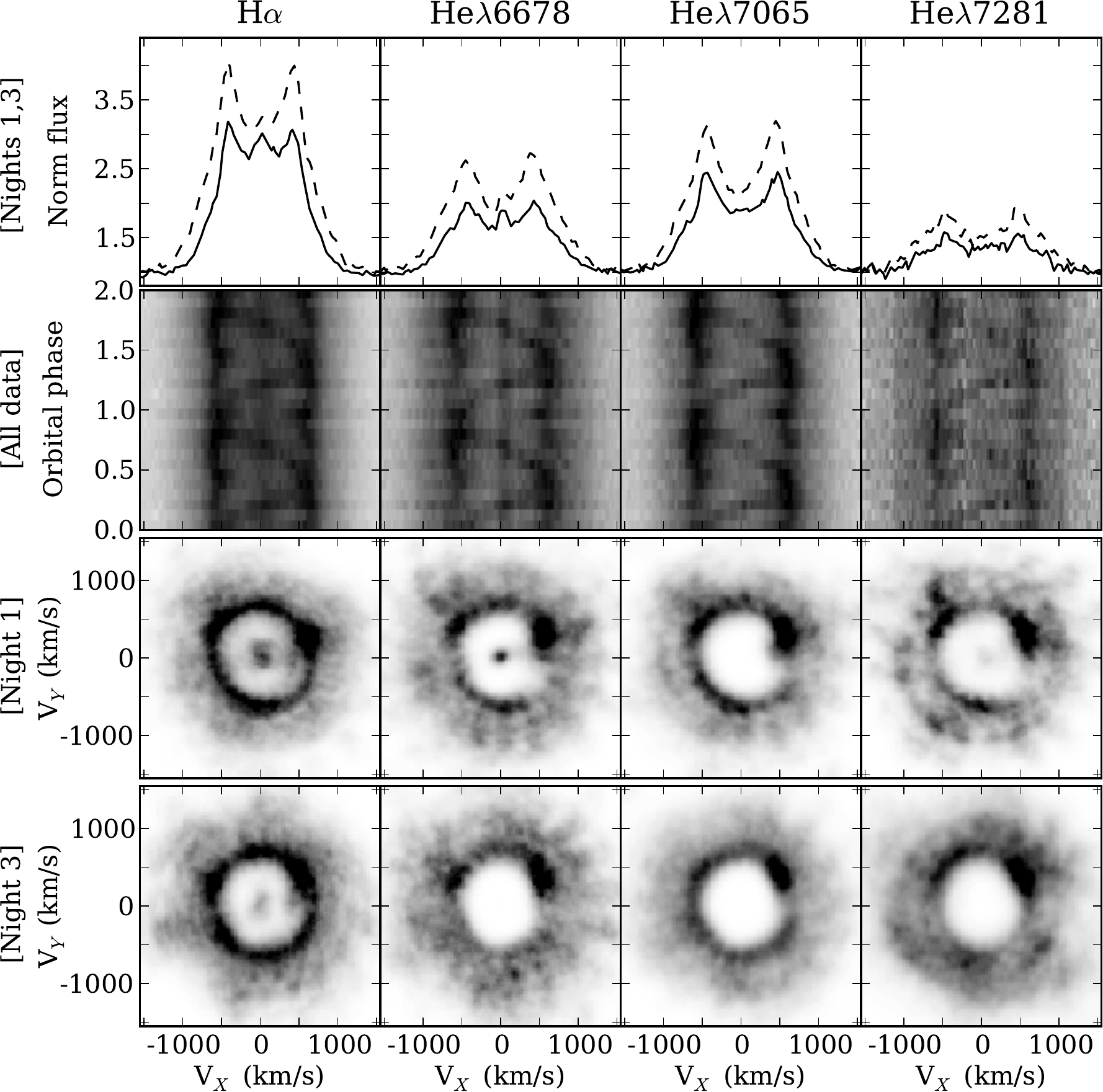}}
\caption{\label{fig:linesdopp} {\it Top row:} Average, continuum normalised emission 
line profiles from the first (solid line) and third (dashed line) nights of observations on VLT/FORS2. 
The double-peaked profile is characteristic of Doppler motion in an accretion disc.
On the first and second night, the \ha\, and \heil\,6678 lines additionally
show a narrow component at the centre of the line. The lines also get broader as they get brighter, 
indicating that the emission originates from a higher velocity region. This is true even if the lines are 
normalised to the line peaks. The vertical axis is the same for all four line profiles, for ease of comparison.
{\it Second row:} Trailed spectra from all three nights, normalised to the continuum and folded on the orbital period. Two orbital cycles are shown for clarity. The panels were scaled individually to the minimum and maximum flux in each panel. 
{\it Third row:} Doppler maps calculated from the first night's spectra. The central spike of the emission line maps to a bright spot at the origin of the map, consistent with the expected velocity position of the white dwarf.
{\it Bottom row:} Doppler maps calculated from the third night's spectra. The white dwarf emission component has disappeared, as it is dominated by the disc emission. The Doppler maps are subject to an unknown rotation compared to the coordinate system such maps are usually displayed in (see text), as the zero phase of the white dwarf is not known. We used BJD$_{0}=2455646.6655$ as in Fig.~\ref{fig:period}.}
\end{figure*}

We show the average line profiles of the four emission lines visible in our FORS2 spectra in Figure~\ref{fig:linesdopp}. Unfortunately, much of the \heil\,5876 emission line is lost from the blue end of the CCD, so this line is not usable for analysis. The lines are broad and display a classic double-peaked profile, revealing the presence of an accretion disc. The \ha\, and \heil 6678 lines additionally have a narrow emission component at the line centre. The \heil 7065 and \heil 7281 lines show no such component. Such a `central spike' is never seen in the spectra of CVs, but many AM\,CVn systems display this feature\footnote{For completeness, we note that the \ha\, emission line in the CV SDSSJ003941.06+005427.5 is triple-peaked 
\citep{southworth10_0039}, but unlike the central spike observed in AM\,CVn stars, it has a large radial velocity amplitude (202~km\,s$^{-1}$), which does not match the expected velocities of the white dwarf, the secondary star, or the accretion disc bright-spot. \citet{southworth10_0039} suggest that it may originate from a coronal loop on a magnetically active secondary star, similar to those that have been observed in outburst spectra of IP\,Peg and SS\,Cyg \citep{steeghs96}.}, e.g. GP\,Com \citep{smak75,marsh99gpcom}, V396\,Hya \citep[CE315,][]{ruiz01} and SDSSJ124058.03-015919.2 \citep{roelofs05}. In these systems the central spikes display radial velocity variations consistent with emission from near the white dwarf \citep{moralesrueda03}. A similar narrow emission line originating from near the white dwarf is seen in some detached post-common-envelope binaries (e.g. LTT\,560, \citealt{tappert11b}, but see also \citealt{tappert11a} and Fig.\,8 of \citealt{gaensicke04}). The white dwarf in LTT\,560 is accreting from the wind of its companion, and the emission is thought to originate in a chromosphere or corona around the white dwarf which forms as a result of the accretion activity.

The central spike in CSS1122-1110 is too weak and its variations too small to reliably measure the radial velocity amplitude with a moving Gaussian model. The trailed spectra in Figure~\ref{fig:linesdopp} show that it is stationary at the line centre to within one pixel (33~km\,s$^{-1}$).
Future observations will greatly benefit from using a shorter exposure time per spectrum to reduce the amount of orbital smearing of the spike, which will allow us to track the small amplitude variations with greater precision.

We observed an increase in line and continuum flux over the three nights of observations. In order to quantify the variability, we shifted the emission lines to their zero velocity position and calculated the average of each night's spectra.  The equivalent widths, as measured from these average spectra, are shown in Table~\ref{tab:lines}. The equivalent widths increase by a factor $\gtrsim1.5$ from the first to the third night. The line flux increased by a factor of 1.6 during the same time, and the continuum, as measured from a line-free part of the spectrum (50\AA\, centered on 6100\AA), brightened by a factor of 1.2. The brightness increase is consistent with the magnitude change measured from the acquisition images (Section~\ref{sec:phot}).  
No periodic signal is detected in the continuum variations. 
We note that, as the system became brighter, the central spike became harder to detect and the lines became slightly broader. The central spike is completely absent from the \heil\,6678 line on the third night (see Figure~\ref{fig:linesdopp}), and barely detected in \ha. We interpret this behaviour as variability originating in the accretion disc, outshining the coronal or chromospheric emission from the white dwarf as it brightens, causing the central spike to disappear from the spectrum.

\begin{table} 
  \centering
  \caption{\label{tab:lines} Line equivalent width variation over the three nights of observations.}
  \begin{tabular}{p{1.5cm} p{0.65cm} p{0.65cm} p{0.65cm}}
     \hline
     ~    & \multicolumn{3}{|c|}{Equivalent width (\AA)}  \\
     \cline{2-4}
     Line    & n1 & n2 & n3 \\
     \hline\hline
     \ha         & 76 & 95 & 113 \\
     \heil\,6678 & 40 & 54 & 69  \\
     \heil\,7065 & 49 & 65 & 78  \\
     \heil\,7281 & 20 & 25 & 35  \\
     \hline 
  \end{tabular}
\end{table}

\subsection{Orbital period}  \label{sec:period}

To measure the radial velocity variation, we used the double Gaussian technique of \citet{schneideryoung80} and \citet{shafter83}, as implemented in {\sc molly}\footnote{{\sc molly} was written by T.R.~Marsh and is available from http://www.warwick.ac.uk/go/trmarsh/software/.}. This measurement is sensitive to the high velocity line wings which originate in the inner disc and is often used as an indication of the radial velocity of the white dwarf \citep[e.g.][]{thorstensen00}.

We used two Gaussian functions of full width half maximum FWHM~=~400~km\,s$^{-1}$ to account for the broad emission lines, and varied the separation between the peaks from 800 to 3000~km\,s$^{-1}$. For each resulting radial velocity curve we calculated the Lomb-Scargle periodogram \citep{lomb76,scargle82} and then folded the radial velocities on the strongest period. The folded radial velocity curve was then fitted with a circular orbit of the form 
\be V(t) = K\;\sin\,[2\pi(t-\mbox{BJD}_0)/$\Porb$] + \gamma\,, \label{eq:sine} \ee
where the reference time BJD$_{0}$ is defined by the blue to red crossing of the measured velocities.

\begin{figure}
\centering
\includegraphics[width=6.5cm]{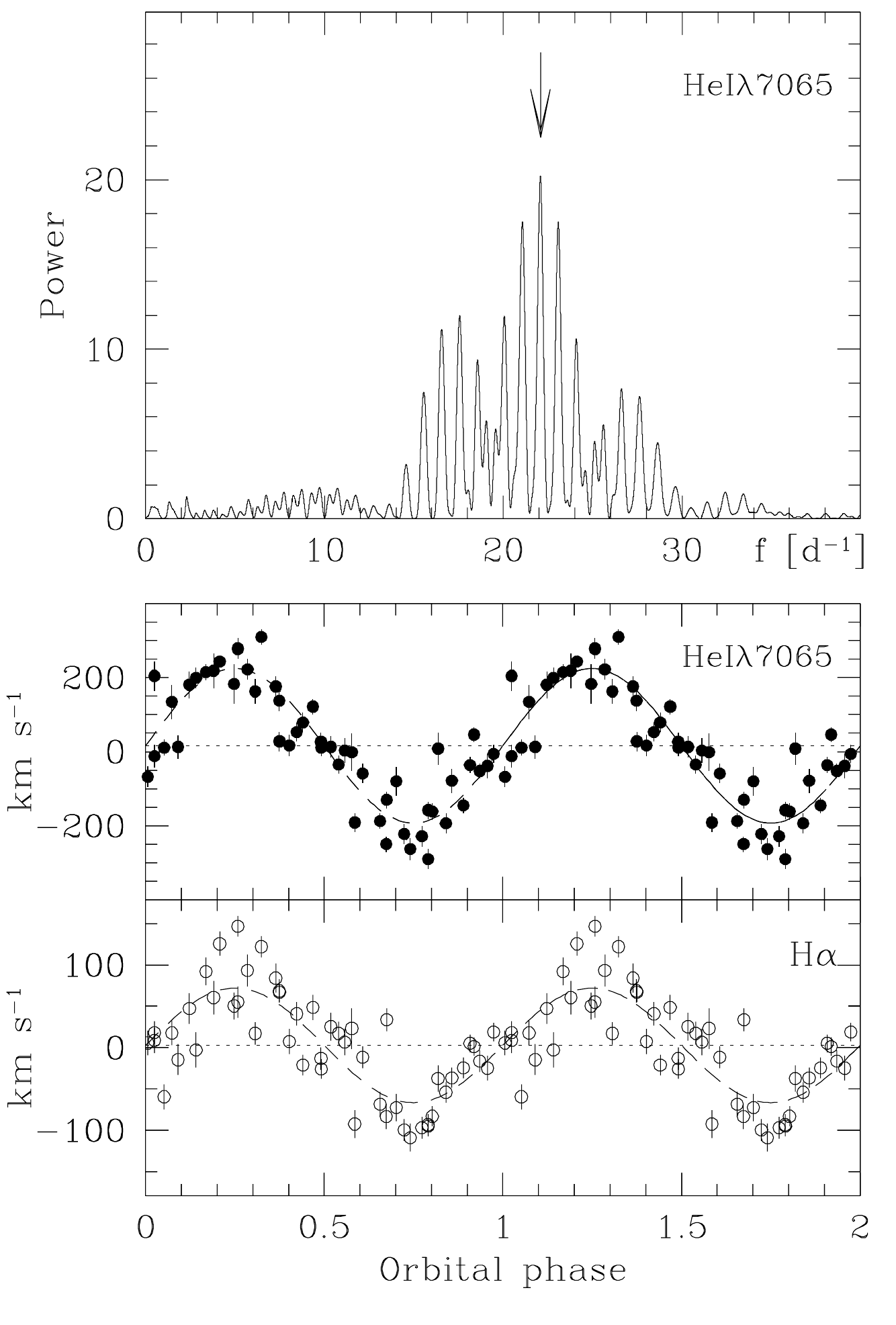}
\caption{{\em Top panel:} Scargle periodogram calculated from the \heil\,7065 radial velocities 
of CSS1122-1110. The adopted orbital frequency is indicated by an arrow. 
{\em Middle panel:} The measured \heil\,7065 radial velocities folded on a period of 65.233
minutes. The dashed line is the best-fit sinusoid to the radial velocities and its parameters
are shown in Table~\ref{tab:prop}. Two phases are plotted for clarity {\em Bottom panel:} 
The \ha\, radial velocities, folded on the same period. Note the smaller amplitude compared to 
the He radial velocity variation.
\label{fig:period}}
\end{figure}

A diagnostic diagram analysis \citep{shafter86} showed that $\sigma/K$ is a minimum for a separation of 1200~km\,s$^{-1}$, where $\sigma$ is the scatter around the fit. This separation also gives a consistent orbital period and reference time from all four emission lines. We show the Lomb-Scargle periodogram calculated from the \heil\,7065 radial velocities in Figure~\ref{fig:period}. The strongest signal occurs at a frequency of 22.075 cycles\,d$^{-1}$, corresponding to a period of 65.233~min. The nearest competing peaks are the $\pm1$~day aliases at 21.058 and 23.064 cycles\,d$^{-1}$. We carried out 10\,000 bootstrap selections of the radial velocity curve, calculated the periodogram of each of these randomly selected subsets and recorded the strongest peak. 98.95 per cent of the radial velocity subsets resulted in 22.074 cycles\,d$^{-1}$ as the strongest signal. 0.6 per cent of the subsets found the 21.058 cycles\,d$^{-1}$ alias marginally stronger, and only 0.45 per cent preferred the alias at 23.064 cycles\,d$^{-1}$. Thus we are confident that the strongest signal is the correct identification of the spectroscopic period of this CV. The \heil\,7065 radial velocity curve, folded on this period, is shown in the middle panel of Figure~\ref{fig:period}. The parameters of the best fit of Equation~\ref{eq:sine}, as derived from the three strongest lines, are listed in Table~\ref{tab:prop}.

The bottom panel of Figure~\ref{fig:period} shows the \ha\, radial velocities folded on the same period. The amplitude of the variations is much smaller than those derived for the helium lines and the curve is highly non-sinusoidal. Its saw-tooth shape suggests that the emission region is non-circular. A non-circular emission region is also supported by the shape of the \ha\, Doppler map (Section~\ref{sec:doppmaps}) in Figure~\ref{fig:linesdopp}. Even though we derive a consistent period from the four available emission lines, the amplitudes of the variation differ by several tens of km\,s$^{-1}$ and in all cases are much larger than expected for the white dwarf in such a compact binary.  This is a clear indication that the phase-dependent modulation of the line wings is not purely a result of the radial velocity of the white dwarf. The high velocity wings of the emission lines contain flux from other bright regions of the disc, e.g. the bright-spot, and cannot be used to track the motion of the white dwarf in this case.

In an effort to isolate the bright-spot emission to measure its velocity amplitude more accurately, we subtract from each spectrum the average spectrum from that night. This removes the constant flux, and leaves only the variable part of the emission. The signal-to-noise ratio is not high enough to fit these residuals in the individual spectra with a moving Gaussian to measure the period, but when folded on the period determined above, the disc bright-spot emission clearly shows up in the Doppler maps and as an S-wave in the trailed spectra. The S-wave is also clearly visible in the continuum normalised spectra shown in the second row of Figure~\ref{fig:linesdopp}. Incidentally, folding the residuals on the either of the 1~day aliases, smears the emission out completely, so that no structure is visible in the trailed spectra. This is further confirmation that we have selected the correct alias as the orbital period. 
Measuring the bright-spot velocity from the phase-binned spectra yields a semi-amplitude of $550\pm27$~km\,s$^{-1}$ for the \heil 7065 line, $575\pm28$~km\,s$^{-1}$ for the \heil 6678 line and $506\pm16$~km\,s$^{-1}$ for \ha. (Table~\ref{tab:prop}).

\begin{table*}
   \caption{Orbital properties as derived from each of the strongest emission lines. \label{tab:prop}}
   \begin{tabular}{lccc}
      \hline
         ~                    & \ha               & \heil\,6678       & \heil\,7065 \\
      \hline\hline 
      \Porb (d)               & $0.04529(2)$      & $0.04532(1)$      & $0.04530(1)$      \\
      BJD$_{0}$               & $2455646.6660(3)$ & $2455646.6663(2)$ & $2455646.6655(2)$ \\
      $K$ (km/s)              & $69.4\pm2.9$      & $285.8\pm5.8$     & $208.4\pm5.1$     \\
      $\gamma$ (km/s)         & $2.8\pm2.0$       & $32.6\pm4.7$      & $16.1\pm3.6$      \\
      Bright-spot velocity (km/s) & $505.8\pm15.7$ & $575.0\pm28.3$   & $549.9\pm26.5$    \\	
   \hline
   \end{tabular}
\end{table*}

\subsection{Binary mass ratio}  \label{sec:q}

Superhumps are characteristic of dwarf novae in superoutburst \citep{whitehurst88,hirose90}. A resonant interaction between the outer  accretion disc and the donor star causes periodic flexing of the disc, which compresses the disc on the side opposite to the donor star and increases the viscous dissipation in this region. The observed brightness modulation is known as superhumps. During the later stages of the outburst, variable dissipation from the bright spot also contributes to the superhump signal, as it sweeps around the edge of the asymmetric disc. See \citet{wood11} for a detailed discussion of the origin of superhumps.  
The dynamical nature of this phenomenon means that there is good reason to expect the superhump period, \Psh,  to be related to the mass ratio $q$ of the binary. Specifically, it is the superhump excess, $\epsilon=$(\Psh--\Porb)/\Porb, which is found to correlate well with $q$ \citep[e.g.][]{patterson05}. This empirical relation is derived from eclipsing dwarf novae, for which the mass ratio and superhump excess can be measured independently. 

There are three similar formulations of the $\epsilon-q$ relation which are commonly used, but unfortunately the low-$\epsilon$ end of the relation is poorly constrained by observations. \citet{patterson05} find $\epsilon = 0.18q + 0.29q^2$ and \citet{kato09} give an updated version as $\epsilon = 0.16q + 0.25q^2$. Both of these formulations assume that $\epsilon=0$ when $q=0$. This is a reasonable assumption, since for a negligible secondary mass, we would expect the tidal interaction with the disc to become negligible as well.
Observations of the only known eclipsing AM\,CVn system, SDSS\,J0926+3624 \citep{copperwheat11}, agree best with the \citet{patterson05} relation, so we favour that version in the calculations below. 
For completeness, we also note the third formulation of the $\epsilon-q$ relationship, as given by \citet{knigge06}, $q = (0.114\pm0.005) + (3.97\pm0.41)\times(\epsilon - 0.025)$. Here the assumption $\epsilon=0=q$ is not made, so the value of $q$ is probably overestimated for this binary. 

Using our spectroscopically determined period and the superhump period measured by \citet{kato10}, we find a small superhump excess in CSS1122-1110, $\epsilon=0.0031\pm0.0007$. This implies an extreme mass ratio, $q=0.017\pm0.004$. 
For the average white dwarf mass found in CVs, $M_1 = (0.83\pm0.23)$\msun\, \citep{zorotovic11}, this implies a very low mass donor, $M_2=(0.014\pm0.005)$\msun.

\subsection{Doppler tomography}  \label{sec:doppmaps}

The second row of Figure~\ref{fig:linesdopp} shows the trailed spectra, centred on each of the emission lines covered by our FORS2 spectra. The spectra from all three nights are included in these trails, each normalised to the continuum and then folded on the orbital period. The central spike is visible as a faint line at the line centres of \ha\, and \heil\,6678, and an S-wave is seen in all four trails. 

We computed corresponding Doppler tomograms \citep{marshhorne88} for each of the emission lines, for each night separately, as well as phase-binned together over the three nights. The increase in the line width as the system gets brigher blurs the map when the spectra from all three nights are included in a single map, so instead we show the maps calculated from the first and the third nights' spectra separately, in the third and fourth row of panels in Figure~\ref{fig:linesdopp}. The maps from the second night's spectra are near-identical to those from the first night. 
All maps were calculated using BJD$_{0}=2455646.6655$ as a phase zero reference time, as derived from the radial velocities of the \heil\,7065 line wings in Section~\ref{sec:period}.  The resulting maps are clearly rotated with respect to the standard orientation usually used in these displays, which has the white dwarf and donor star aligned along the vertical $V_X=0$ axis \citep{marshhorne88}. In that orientation the disc bright-spot appears in the upper left quadrant of the map. This rotation seen here is a result of the unknown phase shift between our choice of BJD$_{0}$ and true zero phase of the white dwarf.
The maps all show a similar structure, with a bright ring corresponding to emission from the accretion disc, a bright-spot on the disc rim (equivalent to the S-wave in the trailed spectra), and a small spot at the origin of the map, which corresponds to the central spike seen in the average \ha\, and \heil\,6678 line profiles. We measured the velocity position of the central spike by fitting it with a two-dimensional Gaussian function. Its velocity is consistent with zero, so we are not able to use it to determine the zero phase orientation of the map. 
We have observed more than one binary orbit per night, so we have also calculated Doppler maps for each orbit individually. From the position of the central spot in these, and the combined maps, we can constrain the radial velocity of the white dwarf to $K_1<16$~km\,s$^{-1}$. On the third night of our observations, when the system was at its brightest, the emission was dominated by the accretion disc. As a result, the central spot disappeared completely from the \heil\,6678 map, and only some weak emission remained at this position in the \ha\, map. 

\begin{figure}
\centering
\rotatebox{0}{\includegraphics[height=7.0cm]{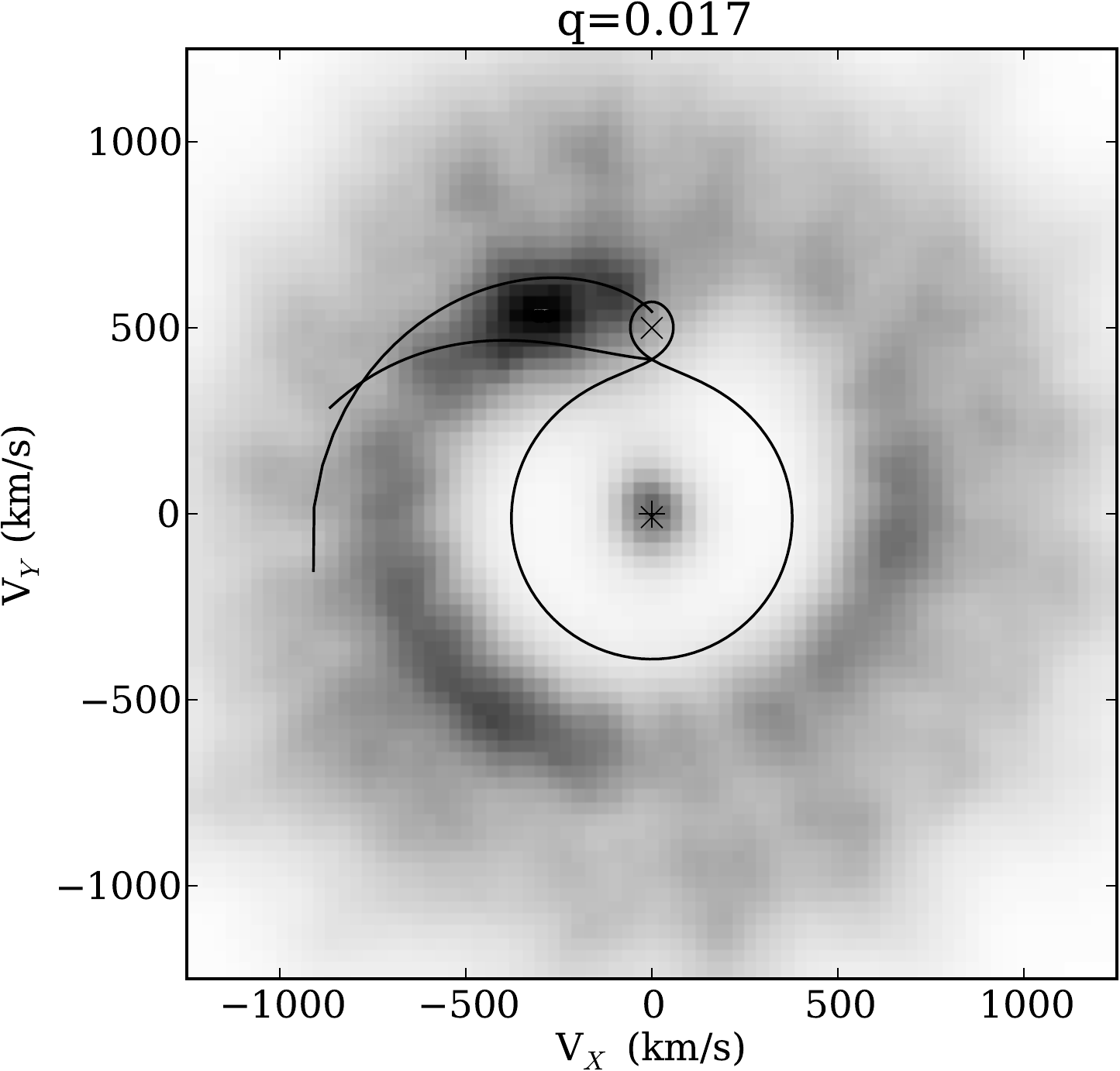}}
\caption{\label{fig:dopproche} \heil\,6678 Doppler map overplotted with the Roche lobes of a $q=0.017$ binary. The assumed velocities of the white dwarf and donor star are labeled with $\times$~symbols and the centre of mass velocity at (0,\,0) with a $+$ symbol. The map was rotated by applying a $-0.25$ phase shift compared to the maps shown in Fig.~\ref{fig:linesdopp}}
\end{figure}

As an illustration, we overplot the \heil\,6678 map with the Roche lobe parameters of a $q=0.017$ binary in Figure~\ref{fig:dopproche}. The assumed velocity positions of the white dwarf, donor star and velocity streams
are also shown. We stress that, since the zero phase of the white dwarf is not known, we have the freedom to rotate the map about its origin, so this combination of parameters is not a unique description of this Doppler map. It only serves as an illustration of the extreme mass ratio $q=0.017$, as determined from the  $\epsilon-q$ relation in the previous section.

\section{Discussion}  \label{sec:discussion}

Three channels are thought to contribute to the formation of AM\,CVn stars. (a) A double-degenerate channel, where a white dwarf accretes from another, helium-rich white dwarf \citep{paczynski67,faulkner72}. This ultra-compact configuration requires a second common envelope phase during the evolution to shorten the period to the observed values, in addition to angular momentum loss by gravitational radiation. 
Potential direct progenitors for this type of binary are the detached double white dwarf systems, in particular those with extremely low mass white dwarf companions \citep{brown11,brown12}.
(b) An alternative outcome of a double common envelope evolution is that the donor star is not fully degenerate when it leaves the common envelope \citep{savonije86,ibentutukov87}. The system starts mass transfer with the donor in the helium burning stage and the donor becomes increasingly degenerate as it evolves towards the period minimum of $\sim10$~minutes. (c) If the initial separation is too large for a second common envelope to form, the system may evolve as a CV. An ultracompact binary may form if the initial mass ratio is such that the donor star undergoes significant nuclear evolution before the onset of mass transfer \citep{podsiadlowski03}. The density of the evolved donor allows the system to reach a much shorter orbital period than is possible for the hydrogen CVs, with the exact value depending on the level of hydrogen depletion in the donor when the mass transfer starts \citep[e.g.][]{sienkiewicz84}. The donor stars in these systems will have some hydrogen left in their envelopes as they evolve towards their period minima, but will become increasingly helium-rich during the evolution. This is the only formation channel in which the donor is predicted to have residual hydrogen in its envelope, so the detection of  hydrogen in the spectrum of an AM\,CVn binary conclusively identifies it as having evolved along this track \citep{podsiadlowski03,nelemans10}. 

The evolved CV formation channel is often dismissed as improbable or unimportant compared to the other channels, since it would require an evolutionary timescale comparable to, or even exceeding, the Hubble time to reach periods as short as those observed in AM\,CVn systems, if angular momentum loss is driven by magnetic braking and gravitational radiation \citep[][Yungelson et al., in prep]{nelemans10}. 
From population synthesis models, \citet{nelemans04} estimate that less than two per cent of AM\,CVn binaries can form this way. However, if the initial mass ratio is $q\gtrsim1$, the binary will undergo a period of thermal timescale mass transfer (TTMT) until the mass ratio decreases to $q\lesssim5/6$, allowing stable mass transfer \citep{tutukov85,pylyser89,schenker02}. The rapid mass loss strips the donor from its hydrogen envelope, exposing the partially evolved core. When the system resumes normal mass transfer and angular momentum loss, it will appear essentially as a normal CV, except that the donor will be evolved compared to other CVs at the same orbital period. Those systems in which most of the hydrogen is lost during the period of TTMT, will be able to evolve to periods well below \Pmin. Since the rapid mass transfer also removes angular momentum at a high rate, it is possible for these systems to evolve to ultrashort periods within a Hubble time. The fraction of CVs with evolved secondaries is expected to be high. \citet{podsiadlowski03} put this figure at ten per cent, and \citet{schenker02} suggest a fraction as high as a third. 

Observations of V485\,Cen \citep[59~min,][]{augusteijn93,augusteijn96} and EI\,Psc \citep[64~min,][]{thorstensen02} support the idea that binaries which have undergone TTMT can evolve to an ultracompact configuration.  The optical spectrum of EI\,Psc shows a donor star with spectral type K4$\pm$2 \citep{thorstensen02}, which is anomalously hot for a main sequence star at such a short orbital period. Its UV spectrum also reveals a low carbon to nitrogen abundance \citep{gaensicke03}, indicating that the accreted material is CNO processed, so the white dwarf is accreting from the exposed, evolved core. Apart from V485\,Cen, EI\,Psc and CSS1122-1110,
there are two more short period binaries which are candidates for AM\,CVn progenitors  
of the `evolved CV' channel: CSS090331:102843-081927 \citep[52~min,][hereafter CSS1028-0819]{woudt12} and CSS111019:233313-155744 \citep[62~min,][hereafter CSS2333-1557]{woudtwarner11_cs2333}. At the moment, little is known about either of these systems. CSS1028-0819 was discovered in outburst by the CRTS and subsequent outburst photometry revealed a period excess of $\epsilon= 0.0530$ \citep{woudt12}. VSNET observers \citep{vsnet11166} reported the outburst spectrum to show both hydrogen and helium lines. No spectroscopic observations of CSS2333-1557 are available yet to determine whether it is an AM\,CVn system, a hydrogen accreting binary or a halo CV\footnote{One more short period hydrogen-accreting binary is known, SDSS\,J150722.30+523039.8 \citep[67~min,][]{littlefair07}. It is excluded from the discussion, as it has been shown to be a CV in the Galactic halo, with a low mass Population II donor star \citep{uthas11}. Its short period stems from the lower atmospheric opacity of such a low metallicity star, which  means that it has a smaller radius and fills its Roche lobe at shorter orbital periods.}.

There are only 42 semi-detached binaries known with periods below \Pmin: 36 AM\,CVn systems, 1 low metallicity halo CV, 3 ultracompact hydrogen accreting binaries and 2 binaries discovered as a result of outburst activity, but still of unknown nature. The three confirmed ultracompact hydrogen accretors (CSS1122-1110, V485\,Cen and EI\,Psc) therefore account for seven per cent of this total. If CSS1028-0819 and CSS2333-1557 are confirmed as hydrogen accreting systems as well, this figure will rise to 12 per cent. Although these figures do not take selection effects into account, such a high fraction of possible progenitors suggests that the evolved CV channel is non-negligible when considering AM\,CVn formation. A statistically complete sample of ultracompact accreting binaries is not yet available. 

\begin{figure}
\centering
\rotatebox{270}{\includegraphics[width=5.5cm]{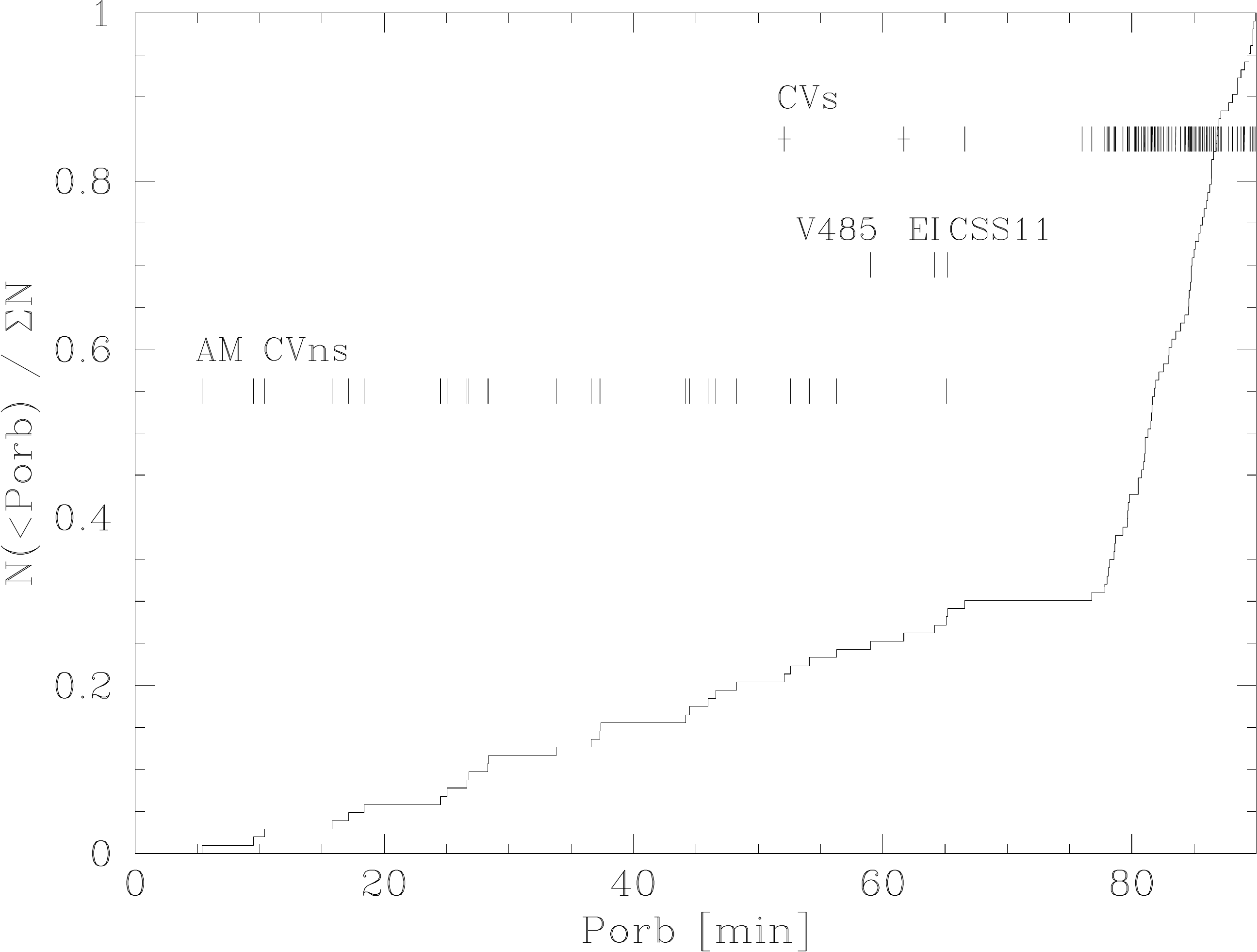}}
\caption{\label{fig:cumulative} Cumulative period distribution of all AM\,CVn stars, CVs with \Porb$<90$~min and short period hydrogen accreting binaries with known orbital periods.  AM\,CVn stars span the period range $5 - 65$ minutes, CVs the range 76 minutes and longer and the ultracompact hydrogen accretors $59 - 65$ minutes. There are no known accreting white dwarf binaries with an orbital period between 67 and 76 minutes. The periods were taken from \citet{solheim10}, \citet{levitan11}, \citet{rkcat}, \citet{gaensicke09}, \citet{augusteijn96}, \citet{thorstensen02}, \citet{littlefair07} and \citet{woudt12}. Individual periods are marked by ticks. The unclassified binaries CSS1028-0819 and CSS2333-1557 are included as CVs, but highlighted by a small horizontal bar. We use an abbreviated notation to indicate the periods of V485\,Cen, EI\,Psc and CSS1122-1110.}
\end{figure}

We show the cumulative period distribution of all accreting white dwarf binaries with known orbital periods in Figure~\ref{fig:cumulative}.  In the period range $59 - 66$ minutes both long period AM\,CVn stars and  short period hydrogen accreting systems are found, but currently there are no known systems with periods between 67 and 76 minutes. This is surprising, because if the critical parameter which determines the minimum period a system can reach is simply the level of hydrogen depletion when the mass transfer starts, there should be no reason not to find binaries in this period range \citep[see e.g.][]{sienkiewicz84}.

Unlike for EI\,Psc and V485\,Cen, the secondary star is not detected in our spectra of CSS1122-1110 (see Figure~\ref{fig:spec}), which means that it must be of later spectral type than the K~star seen in EI\,Psc. The argument for a very low mass, late type donor (Section~\ref{sec:q}) is further supported by the mass ratios of those binaries, which are less extreme than that of CSS1122-1110. For V485\,Cen, we use the \Psh\, measurement of \citet{olech97} to calculate $q=0.13\pm0.01$, and for EI\,Psc, the \Psh\, estimate from \citet{uemura02} gives $q=0.14\pm0.01$. The mass ratio we find for CSS1122-1110 is much more extreme, $q=0.017\pm0.004$. It is however similar to the mass ratios of the long period AM\,CVn stars, e.g. GP\,Com (\Porb=46.6~min, $q=0.018$) and V396\,Hya (\Porb=65.1~min, $q=0.0125$) \citep[both taken from][]{nelemans05}, which suggests that it may be similarly evolved. 
Using a mass-radius relation for a fully degenerate helium star \citep[Eggleton 1986,][]{vr88}, the mass of a such a donor star at an orbital period of 65.233~minutes will be 0.006335\msun, which translates to $q_{\mbox{\tiny min}}=0.0076$. This is even lower than the $q$ measured for CSS1122-1110, indicating that the donor star in CSS1122-1110 is probably semi-degenerate.

The observed properties of CSS1122-1110 strongly suggest that it is an AM\,CVn system forming via the evolved CV channel.
All but the most extreme of AM\,CVn stars which evolve along the evolved CV channel are expected to have hydrogen left in the donor star. The system is expected to become increasingly helium-rich as it evolves to shorter periods, but a small amount of hydrogen could still be left in the donor envelope as it passes the period minimum. The strong (compared to AM\,CVn stars) hydrogen lines in the spectrum of CSS1122-1110  may suggest that is still evolving towards its minimum orbital period. The fact that it shows outbursts also supports this view, since the mass transfer rate must be high enough to trigger the disc thermal instability for outbursts to occur. This is in contrast with GP\,Com and V396\,Hya which have never been observed in outburst, and are believed to have very low mass transfer rates. AM\,CVns showing outbursts are generally found in the 20--40~minute period range \citep[e.g.][]{solheim10}. On the other hand, the very low mass of the donor star could be interpreted as CSS1122-1110 having already evolved past its period minimum \citep[][Table A1]{podsiadlowski03}.

A detailed abundance study may help to determine the evolutionary status of this binary.
The relative abundances depend on a number of factors, such as the hydrogen abundance at the onset of mass transfer, the accretion rate and the temperature structure of the disc. Element abundances are difficult to measure, but \citet{nagel09} successfully reproduced the spectral features of V396\,Hya using a non local thermodynamic equilibrium accretion disc model. The line ratios we measure from the SDSS spectrum of CSS1122-1110 already suggests that the hydrogen content is very low compared to canonical CVs. We measure a flux ratio \heil5876/\ha$=1.49\pm0.04$, compared to \heil5876/\ha$\sim0.2-0.4$ for a typical dwarf novae \citep[e.g.][]{szkody81,thorstensen01}. A more detailed abundance analysis requires a high signal-to-noise spectrum covering a wide spectral range, but a suitable spectrum is not yet available for this system. The relative strengths of CNO species in such a spectrum will further help to distinguish between the different evolutionary scenarios \citep[see][]{nelemans10}.

\section{Conclusion}

Time-resolved optical spectroscopy of the dwarf nova CSS100603:112253-111037 (CSS1122-1110) reveal an orbital period of $65.233\pm0.015$~minutes, well below the CV period minimum. 
As is usual for CVs, the spectra are dominated by \hei\, and Balmer emission lines, but the \hei\, lines are unusually strong, likely a sign of high helium abundance. The \ha\, and \heil\,6678 lines display a narrow emission component, stationary at the line centre to within 16~km\,s$^{-1}$. Using Doppler maps, we can associate this `central spike' with emission from the white dwarf. Such a triple-peaked emission line structure, with a central spike originating from near the white dwarf, is a characteristic only seen in helium-dominated systems, not in CVs. 
We combine our spectroscopically measured orbital period with the superhump period measured by \citet{kato10} to find an extreme mass ratio for this binary, $q=0.017\pm0.004$. Using the average CV white dwarf mass, this implies a very low mass donor star, $M_2 = (0.014\pm0.005)$\msun. Comparing the properties of this ultracompact accreting binary with CVs, AM\,CVn systems and hydrogen binaries below the CV period minimum, we argue that CSS1122-1110 is the first compelling example of an AM\,CVn system forming via the evolved CV channel.


\section*{Acknowledgements}
We thank our referee, M.~Wood, for his suggestions and comments on this paper.
EB, BTG, TRM, DS and CMC acknowledge support from the UK STFC in the form of a Rolling Grant.
This paper is based on observations made with the European Southern Observatory's Very Large Telescope, as part of programme 086.D-0243B.
The CSS survey is funded by the National Aeronautics and Space Administration under Grant No. NNG05GF22G issued through the Science Mission Directorate Near-Earth Objects Observations Program.  The CRTS survey is supported by the U.S.~National Science Foundation under grants AST-0909182.

\bibliographystyle{mn2e}
\bibliography{library3}

\bsp

\label{lastpage}

\end{document}